\documentclass[journal=jpclcd,manuscript=article]{achemso}

\usepackage[version=3]{mhchem} 
\usepackage[dvipsnames]{xcolor}
\usepackage{hyperref}
\usepackage{orcidlink}
\usepackage[flushleft]{threeparttable}
\usepackage{ulem}
\usepackage[dvipsnames]{xcolor}
\usepackage{pdfpages}

\newcommand{\fluenceunit}{$\mu$J/cm$^2$}
\newcommand{\multiple}[1]{$\times$10$^{#1}$}
\newcommand{\densityunit}{cm$^{-3}$ }
\newcommand{\anniunit}{cm$^{-3}$s$^{-1}$}

\author{Yulong~Zheng} 
\affiliation[GT-chem]
{School of Chemistry and Biochemistry, Georgia Institute of Technology, 901 Atlantic Drive, Atlanta GA 30332, United~States}

\author{Rahul~Venkatesh}
\affiliation[GT-chbe]
{School of Chemical and Biomolecular Engineering, Georgia Institute of Technology, 311 Ferst Drive NW, Atlanta GA 30332, United~States}

\author{Esteban~Rojas-Gatjens} 
\affiliation[GT-chem]
{School of Chemistry and Biochemistry, Georgia Institute of Technology, 901 Atlantic Drive, Atlanta GA 30332, United~States}

\author{Elsa~Reichmanis}
\affiliation[Leigh]
{Department of Chemical \& Biomolecular Engineering, Lehigh University, 124 E. Morton Street, Bethlehem PA 18015, United~States}

\author{Carlos~Silva-Acu\~na}
\email{carlos.silva@umontreal.ca}
\affiliation[UdeM]
{Institut Courtois \& D\'epartement de physique, Universit\'e de Montr\'eal, C.P.\ 6128, Succursale centre-ville, Montr\'eal H3C~3J7, Qu\'ebec, Canada}
\alsoaffiliation[GT-chem]
{School of Chemistry and Biochemistry, Georgia Institute of Technology, 901 Atlantic Drive, Atlanta GA 30332, United~States}

\title[An \textsf{achemso} demo]
 {Exciton Bimolecular Annihilation Dynamics in Push-Pull Semiconductor Polymers}

\begin{document}

\begin{tocentry}
\includegraphics[width=4.9cm, height=5.2cm]{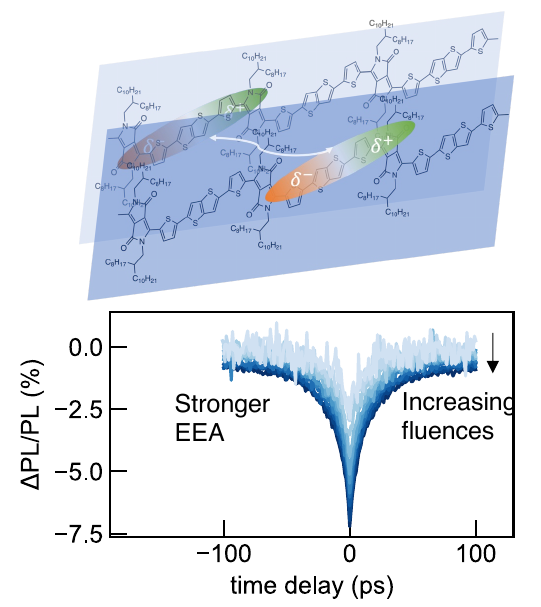}





\end{tocentry}

\newpage
\begin{abstract}
    Exciton-exciton annihilation is a ubiquitous nonlinear dynamical phenomenon in materials hosting Frenkel excitons. In this work, we investigate the nonlinear exciton dynamics of an electron push-pull conjugated polymer by fluence-dependent transient absorption and excitation-correlation photoluminescence spectroscopy, where we can quantitatively show the latter to be a more selective probe of the nonlinear dynamics. Simulations based on a time-independent exciton annihilation model show a decreasing trend for the extracted annihilation rates with excitation fluence. Further investigation of the fluence-dependent transients suggests that the exciton-exciton annihilation bimolecular rates are not constant in time, displaying a $t^{-1/2}$ time dependence, which we rationalize as reflective of one-dimensional exciton diffusion, with a length estimated to be $9 \pm 2$\,nm. In addition, exciton annihilation gives rise to a long-lived species that recombines on a nanosecond timescale. Our conclusions shed broad light onto nonlinear exciton dynamics in push-pull conjugated polymers.
\end{abstract}

\newpage


Frenkel excitons are the primary photoexcitations in conjugated polymers. Following the vertical transitions, excitons experience ultrafast electronic and conformational relaxation to the local minima of the exciton band~\cite{banerji2013sub, banerji2011ultrafast, chang2007intermolecular, fazzi2012ultrafast, tretiak2002conformational}. During this process, a very small percent of the population may dissociate to form polaron pairs in neat conjugated polymer thin films, even if there is no successive two-quantum excitation.\cite{silva2001efficient} Thereafter, excitons can be transported through incoherent hopping.\cite{bredas2004charge, herz2004time} 
When the samples are exposed to sufficiently high laser fluence, the high exciton densities may give rise to singlet exciton-exciton annihilation (EEA). In this work, we probe the EEA dynamics in a conjugated push-pull polymer by comparing transient absorption (TA) and excitation correlation photoluminescence (ECPL) spectroscopic measurements. With a time-independent annihilation model, both trends of the annihilation rates appear to decrease with fluence before a plateau is reached. Previously, the Franck-Condon analysis performed on the absorption lineshape of the same samples prepared from a variety of precursor-solution concentrations reveals an increasing trend of chain backbone order with the viscosity of the precursor solution.\cite{zheng2023chain} In this article, we report that thin films prepared from higher precursor solution concentrations show higher annihilation rates, likely due to short-range Coulombic interactions and/or wave function overlap enhanced by the chain planarization identified previously. Further investigation of the time evolution of exciton density at an early time (20\,ps) in TA indicates that the annihilation rate has a $t^{-1/2}$ dependence, suggesting that exciton diffusion in the push-pull conjugated polymer, DPP-DTT, (poly[2,5-(2-octyldodecyl)-3,6-diketopyrrolopyrrole-\textit{alt}-5,5-(2,5-di(thien-2-yl)thieno-[3,2-b]-thiophene)]) is one-dimensional. In addition to the short-time decay trace, the long-lived tail prevails with increasing pumping fluences, which shows a quadratic dependence, indicating an increasing yield of charges through EEA.

Previously, two mechanisms have been proposed to explain the annihilation process: one is that the annihilation is achieved through F\"orster-type long-range Coulombic interaction.\cite{daniel2007monte} Due to the random spatial distribution of excitons, the ensemble-averaged annihilation rates will decrease with time.\cite{greene1985singlet,forster1949experimentelle, herz2004time} Another model considers the anisotropy of exciton diffusion\cite{tamai2014one, murata2020two} and excitons can only interact when they are in proximity, either through short-range Coulombic interaction or wave function overlap\cite{bredas2004charge, nguyen2000controlling}. In either scenario, the temporal dependence of the annihilation rates reflects on the spatial dependence of the exciton distribution or their motion. Despite the fact of that the pump fluences used in these measurements are orders of magnitude higher than the solar power, the extracted annihilation parameter with the fluence dependence could be theoretically extrapolated to a fluence-independent value, which suggests the ability of intrinsic exciton diffusion. Subsequent to annihilation, one exciton gets deexcited to the ground state while the other is promoted to a higher excited state. While energy relaxation to the low-lying excited state could still occur, the probability of the high-lying excited state dissociating to polaron pairs also increases.\cite{zhu2009charge} Therefore, new long-lived excited species could also be observed with increasing pump fluences.\cite{wang2021intrachain}

The nonlinearity and temporal dependence of EEA processes could distort the monoexponential dynamics on a picosecond time scale in traditional time-resolved measurements, such as transient absorption (TA) and time-resolved photoluminescence (PL).\cite{daniel2003exciton,nguyen2000controlling, lewis2006singlet, shaw2010exciton, gelinas2013recombination, tamai2014one} The mixing between the natural monoexponential decay, EEA, and other linear photophysical processes prohibits the isolation of nonlinear processes from the temporally-resolved signals. In comparison, excitation-correlation (EC) spectroscopy can provide a more selective response to nonlinear dynamics like EEA due to double-amplitude lock-in detection. EC spectroscopy employs two laser beam replicas, each modulated with one chopper at a slightly different frequency.\cite{von1981picosecond, rosen1981time} Therefore, the linear PL from each channel can be acquired when demodulating at each frequency. Furthermore, the nonlinear population mixing arising from EEA between the two beams can also be acquired when the signal is demodulated at the sum of frequencies. Commonly, the EC signals, $\Delta PL/PL$, are demonstrated as a proportion of the nonlinear signal from the sum of nonlinear and linear signals from all three demodulation channels. With the relative arrival time between the two beams controlled by the delay stage, the evolution of the nonlinear dynamics can be further mapped. Although excitation correlation photoluminescence (ECPL) and photocurrent (PC) techniques are not as widely used as TA or time-resolved PL, their applications have always resurfaced with discoveries of new excitonic materials first from a variety of inorganic semiconductors\cite{johnson1988picosecond, chilla1993origin, pau1998}, nanotubes\cite{hirori2006exciton, miyauchi2009femtosecond}, to more recent two-dimensional dichalcogenides\cite{vogt2020ultrafast}, hybrid organic-inorganic perovskites\cite{srimath2016nonlinear, valverde2021nonlinear, perini2022interface, shi20223} due to their sensitivity to nonlinear photophysical responses. Of particular relevance to organic semiconductors, Rojas-Gatjens \textit{et al.} recently investigated the nonlinear PL and PC responses of an organic small-molecule photovoltaic material, where the dominant source of charge carrier generation is ascribed to the EEA process\cite{rojas2023resolving}. Compared to the conjugated homopolymers, conjugated push-pull polymers inherit strong charge-transfer character due to the differences in the electronegativities of the electron-deficient and -sufficient domains, which could have another contribution for the driving force of EEA.\cite{chang2021hj} Here, our work provides new insight into exciton diffusion in conjugated push-pull polymers by comparing the TA and ECPL measurements, experimentally and theoretically, which can be further developed in new optoelectronic systems. 


\begin{figure}
    \includegraphics[width=13cm]{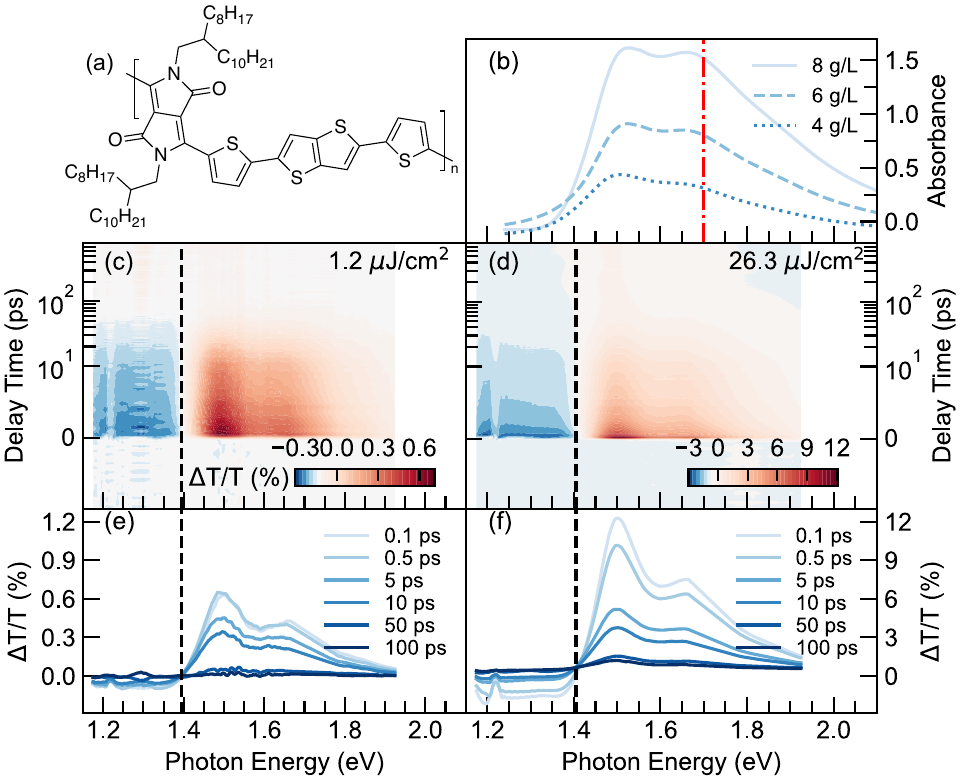}
    \caption{(a) Molecular structure of the repeating unit of DPP-DTT. (b) The absorption spectra of the DPP-DTT thin films prepared from precursors solutions of 4\,(dotted), 6\,(dashed), 8\,(solid) g/L. The red dot-dashed line indicates the pump wavelength used in TA and ECPL measurements.(c) and (d) are transient absorption maps for samples of 8\,g/L, excited by the pump wavelength of (730 nm or 1.70 eV) under low and high fluence, respectively. The 1.2 and 26.3 \fluenceunit corresponds to the excitation density of 3.8\multiple{17} and 8.3\multiple{18} \densityunit (see SI for experimental details). (e) and (f) are the temporal cuts for the spectra. The dash line is a guide for eye to determine the zero cross point.}\label{fig1}
\end{figure}

We focus on a push-pull conjugated polymer, DPP-DTT (Fig.~\ref{fig1}a), following previous ultrafast measurements on this material.~\cite{zheng2023chain} A series of samples prepared from precursor solutions of 4, 6, and 8\,g/L in chlorobenzene were cast using the blade coating technique. The detailed sample preparation process and characterizations are described elsewhere.\cite{venkatesh2021data} The absorption spectra in Fig.~\ref{fig1}b show that the vibronic ratio of 0-0 and 0-1 transition decreases with increasing concentrations, suggesting increasing interchain excitonic interactions.\cite{zheng2023chain} To probe the exciton dynamics, the fluence-dependent TA measurements are first performed under an excitation wavelength of 730\,nm. Here, measurements of the 8\,g/L sample under the lowest and highest fluence are displayed in Fig.~\ref{fig1}c and d, respectively. The other TA measurements with intermediate fluences are also shown in the Fig.~S1 in Supporting Information (SI). Both measurements show similar spectral responses with strong ground state bleaching (GSB) from 1.4-1.9\,eV and photoinduced absorption (PIA) beyond 1.4\,eV. It is worth pointing out that the 2D map of the higher pumping fluence shows a weak, long-lived species, which will be examined in more detail later. The temporal cuts of the spectra are also shown correspondingly in Fig.~\ref{fig1}e and f. A small spectral shift (less than 10\,meV) is noticed between the two fluences, which could be ascribed to the induced electric field under excessive exciton densities \cite{paquin2011charge}. The decay traces are further examined at 750 nm within the GSB region, where the oscillator strengths stem from the 0-0 vibronic Frenkel exciton. We assume that the primary PL and GSB share the same dynamics since only the first excited states are mostly populated. Such assumption allows the following EEA equations to be applicable to both TA and ECPL spectroscopies.

To account for the exciton decay trace, a simple bimolecular exciton-exciton annihilation decay equation reads as,
\begin{equation}
    \frac{dn}{dt} = -\alpha n-\beta n^2,
    \label{eq1}
\end{equation}
where $\alpha$ is the monomolecular exciton decay constant, while $\beta$ denotes the EEA rate constant. It is worth noticing that Eq.~\ref{eq1} assumes that the natural exciton decay and time-independent EEA process are the only two primary pathways for exciton decay, which contribute to the final PL signals, where secondary dynamic processes and excited species could also contribute in reality.\cite{gelinas2011binding, gelinas2013recombination} For example, charge-transfer excitons could be generated either directly\cite{paquin2011charge,roy2017ultrafast, de2016tracking, bakulin2016ultrafast} or through exciton dissociation from a higher energy excited state\cite{silva2001efficient}. Charge recombination could give rise to delayed PL with power-law time dependence.\cite{gelinas2011binding, gelinas2013recombination} Nonetheless, the primary excitation dominates the majority of the PL signals and the EEA mechanism should serve as the simplest quantitative case study. The equation has an analytical expression,
\begin{equation}
    n(t) = \frac{\alpha n_0}{(\alpha + n_0\beta)e^{\alpha t}-\beta n_0}.
    \label{eq2}
\end{equation}
Eq.\ref{eq2} can be further linearized into~\cite{lewis2006singlet, shaw2008exciton},
\begin{equation}
    \frac{1}{n(t)} = \left(\frac{1}{n_0}+\frac{\beta}{\alpha}\right)e^{\alpha t}-\frac{\beta}{\alpha},
    \label{eq3}
\end{equation}
the initial excitation density is given as $n_0$ upon excitation. A quick examination of Eq.~\ref{eq3} shows that the inverse of the excitation density should have a negative intercept. 

\begin{figure}
    \includegraphics[width=0.5\textwidth]{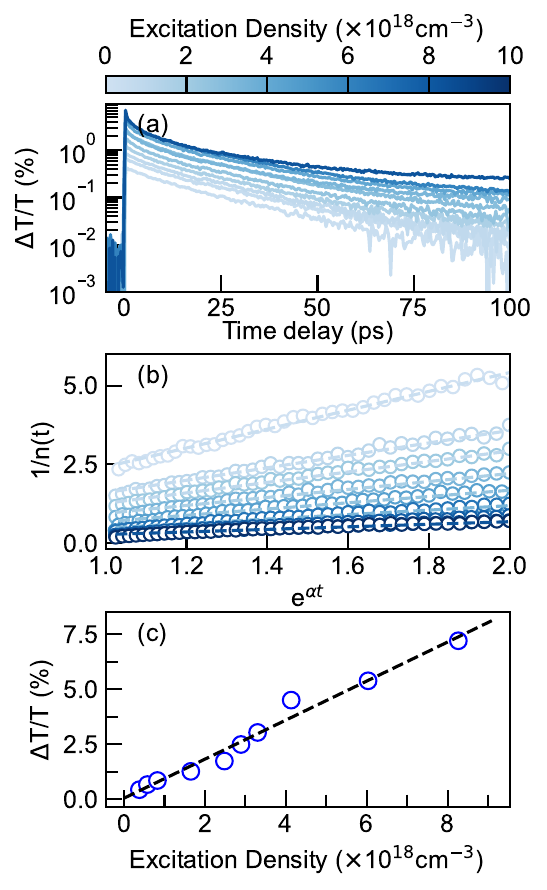}
    \caption{(a) TA Decays at 750 nm with varying excitation densities. With increasing excitation densities, a faster decay is observed within the first few picoseconds. (b) Linearized TA decays (white open circles) fit with Eq.~\ref{eq3} (dashed straight lines). The first 20 ps is chosen and converted for the exponential \textit{x}-axis. (c) The dependence of initial differential transmission (open circles) on excitation densities. The black dash line fits the linear relationship with a slope of 0.889($\pm$0.047) and intercept of 0.042($\pm$0.183).}
    \label{fig2}
\end{figure}

To extract the bimolecular annihilation rate, $\beta$ in the form of Eq.~\ref{eq3}, the fluence-dependent temporal cuts at 750\,nm from TA are plotted in Fig.~\ref{fig2}a. At relatively low fluences, the log-scale differential transmission traces show a mostly linear dependence on delay time, while within 20\,ps, the nonlinear decaying component due to EEA becomes more prevalent. The monoexponential decay rate, $\alpha$ is fixed at 0.053\,ps$^{-1}$ as exctrated from an exponential fit, excited by the lowest pump fluence\,(1.2\,\fluenceunit), which is assumed to be in the regime of dominated monoexponential decay. Therefore, $\beta$ can be acquired by solving the slope and intercept of the linear fit together, as shown in Fig.~\ref{fig2}b. Before moving on to discussing the acquired annihilation rates, it is worth pointing out that the extraction of the annihilation rates relies on the assumption that the initial differential signal is attributed to a single-step pumping excitation. As shown by Silva \textit{et al.},\cite{silva2001efficient} two-step excitation originating from the leading and trailing edge of a single pulse could also lead to nonlinear decaying dynamics in TA, which mixes with the EEA source. However, as shown in Fig.~\ref{fig2}c, the differential transmission signals at time zero not only have a linear dependence on the excitation density, but also have an almost 0 y-intercept (0.042\,$\pm$\,0.183), which excludes the possibility of two-step excitation. Based on Eq.~\ref{eq3}, the annihilation rates can be readily calculated since $\alpha$ is known and $n_0$ can be estimated with laser fluence, film thickness, and the absorption coefficients. 


\begin{figure}
    \includegraphics[width=0.5\textwidth]{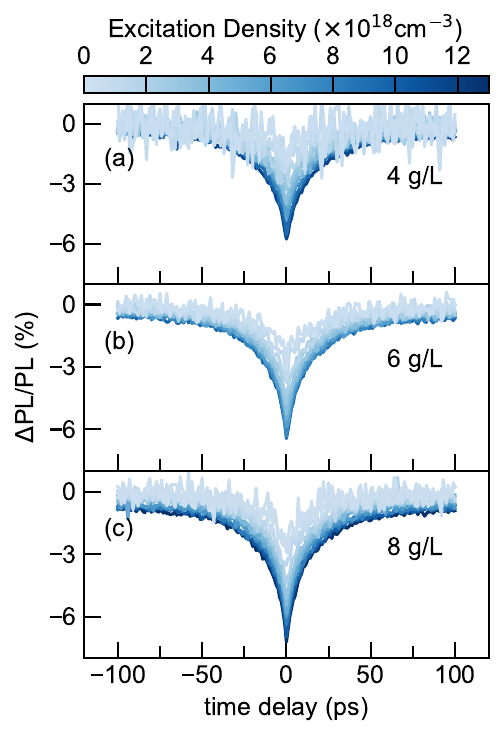}
    \caption{ECPL nonlinear response profiles excited at 730 nm pump for samples prepared from 4~(a), 6~(b) and 8\,g/L~(c) precursor solutions. The PL signals were filtered to collect the wavelength range of 750-1100\,nm. The measurements are performed under a variation of excitation densities coded by the colorbar scale.}
    \label{fig3}
\end{figure}

The annihilation rates acquired from TA measurements can be further compared to those of their ECPL counterparts. Prior to that, we resort to deriving an annihilation-based model in describing the ECPL signal profiles. Previous work revealed that with samples prepared from higher concentration solutions, polymer interchain excitonic interaction increases, as well as the chain backbone planarity.\cite{venkatesh2021data, zheng2023chain} Both factors might contribute to a distinct strength of exciton-exciton interaction. With the aforementioned ECPL working principle, all ECPL profiles measured on DPP-DTT thin films of different precursor concentrations demonstrate a negative signal and diminish with delayed times between the two pulses, as shown in Fig~\ref{fig3}. A detailed description of the ECPL setup can be found in the SI. The overall negative signals reflect EEA as an efficient linear PL quenching pathway, while the decaying nonlinear signals originate from the less temporal overlap between the two pulses, thus less sufficient population mixing. 
To analyze the results quantitatively, we further implement Eq.~\ref{eq2} based on lock-in detection, which essentially gives rise to a time-integrated signal,
\begin{equation}
    PL_{ind} = \int_0^{\infty}n(t)dt
    = 1/\beta \{ln[(1+\gamma)]\},\\
    \label{eq4}
\end{equation}
where $\gamma$ is a unitless parameter defined as, $\gamma\equiv\frac{\displaystyle n_0\beta}{\displaystyle \alpha}$. Considering the monoexponential decay is constant, the product of the initial excitation density and annihilation rate, thus $\gamma$, is a measure of the strength of the EEA process. On the other hand, the nonlinear signal demodulated at the sum of the chopping frequencies can be toggled through the delay time between the two beams. The varying delay times impact the nonlinear signal in the way that excitons generated from the first pulse will decay until the second pulse comes in. Thereafter, the total amount of excitons should be given as the sum of the residual from the first decay and the newly-generated amount,
\begin{equation}
\begin{split}
    PL_{sum} = \int_0^{\tau}n(t_1)dt_1+\int_0^\infty n(t_2)dt_2\\
    = 1/\beta \{ln[(1+\gamma)^2-\gamma^2 e^{-\alpha\tau}]\}.
    \label{eq5}
\end{split}
\end{equation}
Eventually, the experimentally meaningful equation can be given as,
\begin{equation}
    \Delta PL(\tau)/PL = 1-\frac{2\ln(1+\gamma)}{\ln[(1+\gamma)^2-\gamma^2e^{-\alpha\tau}]}.
    \label{eq6}
\end{equation}
One extreme scenario can be readily inspected: when the time delay $\tau$ approaches infinity, Eq.~\ref{eq6} will give 0, indicating null nonlinear PL, which is expected as the long intervals between the two pulses prohibit the generation of the cross term. As indicated earlier, ECPL should be more selective in separating nonlinear signals than TA. This can be readily seen if we assume no annihilation, suggesting that the excitation should be completely monoexponential. It then can be shown that PL$_{sum}$ is simply double PL$_{ind}$ which is $\frac{\displaystyle n_0}{\displaystyle \alpha}$. Therefore, Eq.~\ref{eq6} will yield 0, which rigorously shows that linear dynamics alone would not give ECPL signals.

\begin{figure}
    \includegraphics[width=0.5\textwidth]{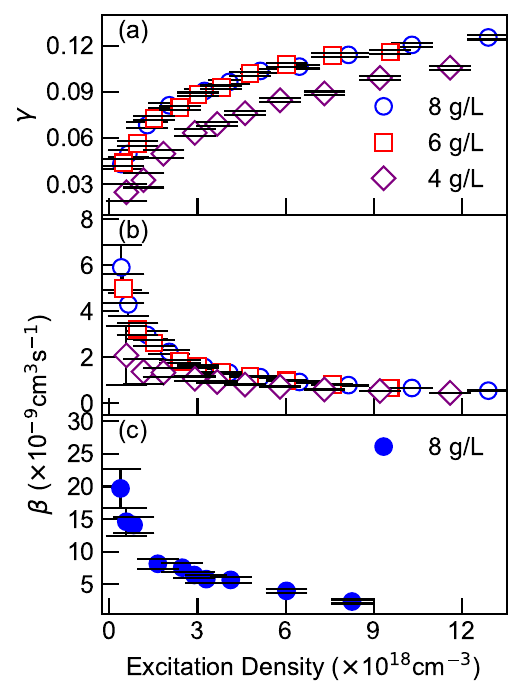}
    \caption{The excitation-density dependence in DPP-DTT thin films prepared from 4, 6 and 8 g/L solutions of (a) unitless parameter, gamma, (b) EEA rates, $\beta$, acquired by fitting ECPL profiles using Eq.\ref{eq4} compared with (c) EEA rates acquired from Figure \ref{fig3}b measured from TA.}
    \label{fig4}
\end{figure}

The complete simulation results are shown in Fig.~S3-5, which demonstrate excellent consistency with the experimental results. The extracted $\gamma$ with increasing excitation densities implies stronger EEA effects as expected (Fig.~\ref{fig4}a). Interestingly, the $\gamma$s acquired from the sample of 4\,g/L are notably lower than those prepared from higher precursor concentrations. Furthermore, simulations based on Eq.~\ref{eq6} yield annihilation rates on the order of magnitude of 10$^{-9}$\,\anniunit (Fig.~\ref{fig4}b). Meanwhile, the annihilation rates extracted from TA also show a decreasing trend with excitation density even with overall higher $\beta$s as shown in Fig.~\ref{fig4}c. Indeed, annihilation rates acquired from time-integrated measurements are frequently shown to be lower compared to the parameters extracted from their time-resolved counterparts for the same type of conjugated polymer.\cite{stevens2001exciton, riley2022quasi} Such difference might be partially ascribed to integrating long-lived PL signals that originate from polaron pair recombination and/or triplet-triplet annihilation.\cite{gelinas2013recombination} Those long-lived PL signals compensate for the PL quenching by EEA in that annihilation rates are underestimated with higher pumping fluences.
Except for slight differences in the absolute values of $\beta$, the annihilation rates show a consistent asymptotic decreasing trend. It is worth mentioning that decreasing annihilation rates are not uncommonly observed. Previous literature ascribed the origins to either excitons generated within the EEA radius annihilating rapidly or excitons with a shorter effective lifetime under higher densities.\cite{lewis2006singlet, riley2022quasi} Nevertheless, excitons generated within the annihilation radius should not be rare even under low excitation fluences as the interaction radius is calculated as an ensemble average. On the other hand, the effective monomolecular lifetime would shorten due to stimulated emission or excited state absorption with fluences, their variations are much smaller in contrast to the multiple times change of $\gamma$ (See Figure~S6 in SI). Alternatively, it is worth pointing out that the annihilation rate could be a time-dependent value, especially in the early stage.\cite{greene1985singlet} Previous publications indicate that such dependence originates from the dimensionality of exciton diffusion, where not only isotropic but also one- and two-dimensional diffusion have been identified in different semiconductor polymers, which might be accountable for the decreasing trend for the annihilation rates with fluences.\cite{tamai2014one, murata2020two, shaw2008exciton} 

\begin{figure}
    \includegraphics[width=0.8\textwidth]{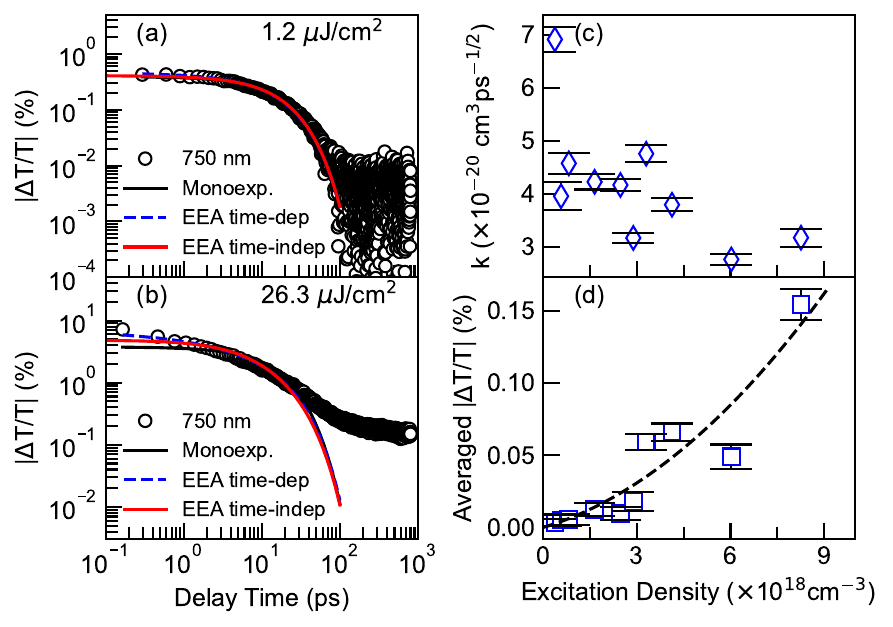}
    \captionof{figure}{(a) and (b) are temporal decays under low and high fluences, respectively. The early-time decays are fitted with monoexponential decay (black solid line), time-independent EEA model(red solid) and time-dependent EEA model (blue dashed line). (c) Dependence of diffusion constants on excitation densities using Eq.~\ref{eq5}. (d) time-averaged differential transmission at 800\,ps with respect to excitation densities. The black dash line is the quadratic fit with the y-intercept set as 0.}
    \label{fig5}
\end{figure}

The exciton annihilation rate could have a $t^{-1/2}$ time dependence due to either the spatial distribution of excitons, which annihilate through long-range Coulombic interaction or one-dimensional diffusion-limited annihilation. In either scenario, the time-dependent annihilation model (Eq.~\ref{eq2}) could be reformulated as\cite{tamai2015exciton},
\begin{align}
    n(t) = \frac{n_0 e^{-\alpha t}}{1+\frac{n_0\pi k}{\sqrt{\alpha}}erf(\sqrt{\alpha t})},
    \label{eq7}
\end{align}
where $k\equiv\beta(t)\times\sqrt{t}$ so that the newly-defined annihilation rate, $k$, can now be simply described as a time-independent term and $erf$ is the error function. For a better comparison, all simulations based on monoexponential, time-independent, and time-dependent models are shown in both the lowest and highest TA decay traces in Fig.~\ref{fig5}a and b, respectively. Under the lowest pumping fluence, all three models fit the dynamics closely indicating that the dynamics at low pump fluence is dominated by monoexponential decay with minor impact from EEA. However, under high pump fluence, a small deviation becomes clear in the early delay times (first 2\,ps) when comparing the time-dependent annihilation model with the other two; the first kind fits the experimental result best till 30\,ps. Calculation of the new annihilation constants, $k$, gives a consistent value of 4\,$\pm$\,1.1\multiple{-14}\,cm$^3$s$^{-1/2}$ as shown in Fig.~\ref{fig5}c. One large outlier can be readily distinguished at the lowest fluence case since the additional annihilation term could be overfitting. Therefore, we suggest that EEA could be a time-dependent process in DPP-DTT.

Another distinct feature is the drastic offset between all simulations and the experimental decay trace beyond 50\,ps at the highest fluence (Fig.~\ref{fig5}b). Furthermore, the long-lived tail no longer follows an exponential decay. To avoid data fluctuation at low signal-to-noise ratio, especially in the low fluence case, 20\,points around 800\,ps are averaged for each excitation density. The eventual signal at long-time delay (LTD) dependence on the excitation density is demonstrated in Fig.~\ref{fig5}d, where a quadratic dependence is observed. The corresponding equation is given by,
\begin{equation}
    |\frac{\Delta T}{T}|_{LTD} =0.00129\,n_0^2+0.0064\,n_0,
    \label{eq8}
\end{equation}
where the y-intercept is set as 0 since no excited species should exist without a pump laser. The long-lived excited species likely originate from the polaron pair, and the quadratic dependence suggests the EEA as the source.\cite{silva2001efficient, wang2021intrachain} Furthermore, since Eq.~\ref{eq8} also has a linear dependence on excitation density, it also suggests that a certain amount of excitons have experienced direct dissociation. Considering the single-step exciton generation from Fig.~\ref{fig5}c, the quantum yield of the polaron pairs due to direct dissociation is estimated to be 0.7\%. This value is significantly lower than a few other conjugated polymer systems, where a quantum yield of 10\% is estimated within the first 150\,fs.\cite{silva2001efficient} One possibility could be that the quantum yield is estimated at a fairly long time delay, where a large proportion has already decayed, leading to an inaccurate estimate.


In this work, we integrate and compare the parameters acquired from both the TA and ECPL measurements based on the exciton-exciton annihilation model. As mentioned earlier, exciton-exciton annihilation can possibly be achieved by two different mechanisms, either through diffusion-limited exciton collision or direct long-range Coulombic interaction. There exists the possibility of EEA suffered from long-range Coulombic interaction, assuming that the time dependence of the EEA rates originates from a spatial ensemble average of exciton interaction. However, in previous work, we showed that the exciton becomes more delocalized with increasing precursors' concentration.\cite{zheng2023chain} As the exciton becomes more delocalized, the transition dipole moments would weaken. The long-range Coulombic interaction would deviate from the dipole approximation to multipole approximation (e.g. quadrupolar interactions), leading to reduced EEA. In addition, it is commonly agreed that incoherent exciton hopping achieved through such F\"orster-type long-range interaction requires sufficient spectral overlap between the absorption and PL. For DPP-DTT, the Stokes shift increased from 130 to 180 meV with increasing precursor concentration\cite{zheng2023chain}, supposedly leading to weaker EEA. Nevertheless, the opposite trend is observed, which suggests that exciton diffusion and collision might also play an important role; EEA might involve short-range interactions through either Coulombic or wave function overlap. Recently, Tempelaar \textit{et al.} calculated the exciton annihilation rates theoretically, assuming that excitons interact through resonant Coulombic coupling.\cite{tempelaar2017exciton} The annihilation rates are found to decrease with decreasing exciton densities, which is the opposite of the trend shown in Fig~\ref{fig4}. Such evidence suggests that the annihilation between excitons through long-range interaction might not be the mechanism considered here. 

It is worth mentioning that long-lived tails have been widely observed in conjugated polymers with a variety of possibilities for their origins.\cite{denton1999optical, mcbranch1999signatures, stevens2001exciton, herz2004time, gelinas2011binding,  gelinas2013recombination} Interchain polaron pairs have been previously identified to be mediated by lattice defects with a linear dependence on pump fluence.\cite{mcbranch1999signatures} Similar behavior might be expected for homocoupling defects due to the synthesis of DPP-based copolymers, giving rise to an unexpected lower-energy shoulder in the absorption spectra,\cite{hendriks2014homocoupling} which is nevertheless not observed in the absorption spectra of this series of samples as shown in Figure~\ref{fig1}b. Considering the quadratic dependence on the pump fluence, both possibilities can be safely excluded. Another source of the long-lived tails might be from the singlet fission of free triplet exciton and/or triplet-triplet exciton pair formation.\cite{huynh2017transient, musser2013activated} In this work, we did not observe distinct feature that can be assigned undoubtedly as triplet excitons. Besides, the triplet-exciton dependence of the fluence should also be linear since only one excited chromophore is involved in the singlet fission process. Therefore, we assign the long-lived tail as observed in this work to the polaron pairs through EEA process, to our best knowledge.

Using the one-dimensional diffusion model, the diffusion coefficients, $D$, can be calculated based on their relation to $k$,\cite{daniel2003exciton}
\begin{equation}
    k = 2\sqrt{2\pi D}R^2,
\end{equation}
where the annihilation radius, $R$, in the diffusion limit, is normally estimated as the lamellar layer distance, $d_{100}$, as extracted from the in-plane profile of grazing incidence wide-angle X-ray scattering.\cite{shaw2008exciton, shaw2010exciton} In DPP-DTT, it is found to be around 2\,nm.\cite{venkatesh2023overlap} Therefore, the diffusion coefficient, $D$, is estimated to be 4\,$\pm$\,2\,nm$^2$ps$^{-1}$ and the diffusion length is given as, $L=\sqrt{D/\alpha}$, which is 9\,$\pm$\,2\,nm. Both values are in good agreement with results found for other conjugated polymers.\cite{tamai2014one, daniel2007monte, riley2022quasi} 

\begin{table}
\begin{threeparttable}
\caption{Comparison of the diffusion lengths acquired from time-dependent and time-independent EEA model acquired from ECPL and TA measurements.}
\begin{tabular}{cccccc}
\hline
\hline
Conc. (g/L)  & 4          & 6          &            & 8              &              \\ \cline{4-6} 
Technique    & ECPL       & ECPL       & ECPL       & Time-indep. TA & Time-dep. TA \\ \hline
$L$ (nm) at $n_{0,l}^a$ & 0.9\,$\pm$\,0.7   & 1.4\,$\pm$\,0.5   & 1.3\,$\pm$\,0.6   & 2.6\,$\pm$\,1.0       & 8.2\,$\pm$\,0.5$^{c}$   \\
$L$ (nm) at $n_{0,h}^b$  & 0.42\,$\pm$\,0.08 & 0.49\,$\pm$\,0.08 & 0.48\,$\pm$\,0.08 & 0.9\,$\pm$\,0.3       & 6.6\,$\pm$\,0.4     \\ \hline
\hline
\end{tabular}\label{tab1}
\begin{tablenotes}
        \small
        \item $^{a, b}$ $n_{0,l}, n_{0,h}$ denote initial excitation density at lowest and highest pump fluence, respectively.
        \item $^c$ value obtained for the second highest excitation density as shown in Figure~\ref{fig5}c. The first point is ignored for its obvious deviation.
      \end{tablenotes}
\end{threeparttable}
\end{table}
To compare the results with the diffusion lengths acquired from the time-independent model, the results are summarized in Table~\ref{tab1}. The diffusion lengths acquired from the time-independent EEA model based on three-dimensional isotropic diffusion\cite{tamai2015exciton}, whether from ECPL or TA, have much smaller values than those from the time-dependent model (5-10 times smaller). Such deviation is inherited from neglecting the dimensionality of exciton diffusion. It can be simply understood as the length of one-dimensional chain will be significantly reduced when 'simulating' it into the radius of a three-dimensional sphere, considering the same volume. In addition, the diffusion lengths derived from the same time-independent EEA model differ by one time, comparing the ECPL and TA measurements. The slight difference could be due to the incorporation of the long-lived emission in ECPL measurements as discussed earlier. Last but not least, the diffusion lengths acquired for the sample of 6 and 8\,g/L are higher than those of lower concentration samples, as the diffusion is aided by the short-range interaction supported by the enhanced chain backbone order.

It is worth mentioning that in our current ECPL analysis, we igonred the contribution from stimulated emission and/or excited state reabsorption from the prompt PL followed by the first pump. Although it can be easily compensated for by loosing the constraint on the monoexponential decay constant, $\alpha$, but its contribution should be investigated rigorously which is outside the scope of this work. In addition, the complicated Eq.~\ref{eq7} obviously prohibits us from getting a simple analytical model for ECPL measurement as was possible with its time-independent counterpart. However, numerical methods such as Genetic Algorithm might be one of the options for achieving a universally applicable model for extracting both monomolecular and annihilation rate constants, which can be further employed in other systems with even more complicated dynamics.\cite{gelinas2013recombination} 

In conclusion, we examine the dynamics of exciton-exciton annihilation in a specific push-pull polymer and compare the experimental and simulation results obtained from transient absorption and excitation correlation spectroscopy. Using the time-independent annihilation model, both measurements yield a decreasing annihilation rate trend with increasing fluence until reaching a plateau. Thin films deposited from higher precursor solution concentrations exhibit higher annihilation rates, likely due to stronger short-range Coulombic interactions or wave function overlap between excitons. By analyzing the time evolution of exciton density at an early stage (20\,ps) in transient absorption, we find that the annihilation rate follows a $t^{-1/2}$ dependence, suggesting one-dimensional exciton diffusion along the chain in DPP-DTT. The one-dimensional diffusion length is estimated to be 9\,nm, which is in good agreement with a variety of other conjugated polymers. Additionally, besides the rapid decay, there is a long-lived tail that becomes more prominent as pumping fluences increase. This tail demonstrates a quadratic dependence, indicating an increasing yield of charges through exciton-exciton annihilation. Our work rigorously shows the application of the ECPL technique in conjugated polymers and a further reach to the wider semiconductor research field.
\begin{acknowledgement}

E.R., R.V. and Y.Z. appreciate support associated with National Science Foundation Grant No. 1922111, DMREF: Collaborative Research: Achieving Multicomponent Active Materials through Synergistic Combinatorial, Informatics-enabled Materials Discovery. E.R also acknowledges support from Carl Robert Anderson Chair funds at Lehigh University. C.S.A appreciates support associated from the National Science Foundation (Grant DMR-1729737). C.S.A also acknowledges the Canada Excellence Research Chair in Light-Matter Interactions in Photonic Materials, and a Courtois Research Chair.

\end{acknowledgement}

\begin{suppinfo}

See the Supplementary Information for the experimental methods for TA and ECPL and their associated measurements and fits under varying fluences.

\end{suppinfo}

\providecommand{\latin}[1]{#1}
\makeatletter
\providecommand{\doi}
  {\begingroup\let\do\@makeother\dospecials
  \catcode`\{=1 \catcode`\}=2 \doi@aux}
\providecommand{\doi@aux}[1]{\endgroup\texttt{#1}}
\makeatother
\providecommand*\mcitethebibliography{\thebibliography}
\csname @ifundefined\endcsname{endmcitethebibliography}  {\let\endmcitethebibliography\endthebibliography}{}

\newpage


\section*{Supplementary material}

# Supporting Information: Exciton Bimolecular Annihilation Dynamics in Push-Pull Semiconductor Polymers

Yulong Zheng[1], Rahul Venkatesh[2], Esteban Rojas-Gatjens[1], Elsa Reichmanis[3], Carlos Silva-Acuña[‡1,4]

[1] *School of Chemistry and Biochemistry, Georgia Institute of Technology, 901 Atlantic Drive, Atlanta, Georgia 30332, USA.*
[2] *School of Chemical and Biomolecular Engineering, Georgia Institute of Technology, 311 Ferst Drive NW, Atlanta GA 30332, United States.*
[3] *Department of Chemical & Biomolecular Engineering, Lehigh University, 124 E. Morton Street, Bethlehem PA 18015, United States.*
[4] *Institut Courtois & Département de physique, Université de Montréal, C.P. 6128, Succursale centre-ville, Montréal H3C 3J7, Québec, Canada.*
E-mail: carlos.silva@umontreal.ca

## Contents

# 1 Transient Absorption Measurements

## 1.1 Experimental Methods

Transient absorption measurements were conducted using an ultrafast laser system, specifically the Pharos Model PH1-20-02-10 by Light Conversion. Tunable wavelengths were achieved by utilizing a laser fundamental with a wavelength of 1030 nm at a repetition rate of 100 kHz. The integrated transient absorption was assessed within a commercial setup, the Light Conversion Hera system. For the pump, wavelengths spanning from 360 to 2600 nm could be adjusted by directing the 10W laser output into a commercial optical parametric amplifier known as Orpheus (manufactured by Light Conversion, Lithuania). On the other hand, the probe beam was generated by sending 2W of power through a sapphire crystal to produce a single-filament white-light continuum within the spectral range of 490-1060 nm. Following this, the probe beam, after passing through the sample, was collected using an imaging spectrograph (Shamrock 193i, Andor Technology Ltd., U.K.) combined with a multichannel detector boasting 256 pixels and covering a wavelength range from 200 to 1100 nm. The pump was selected at 730 nm while the decay traces were monitored at 750 nm. All the samples were measured in a homemade vacuum chamber at ambient temperature.

## 1.2 Experimental Results and Simulations

The transient absorption measurements pumped from low to high fluences for the thin film prepared from 8 g/L precursor solutions are shown in Fig. S1. The two extreme cases (1.2 and 26.3 $\mu$J/cm$^2$) are shown in the main article. All measurements show similar spectral features with ground bleach signals from 1.4 to 1.9 eV, while the photoinduced absorption extends below 1.4 eV.

The temporal decays at varying fluences observed at 750 nm are shown in Fig. S2. It can be clearly observed that with increasing fluences, the tails beyond 300 ps start to deviate from all three models (monoexponential, time-independent and time-dependent annihilation).

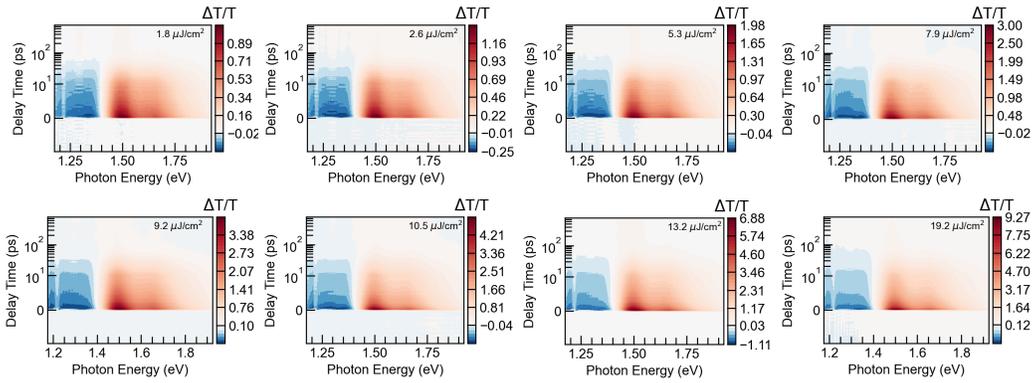

**Figure S1:** transient absorption maps of the 8 g/L thin film sample from the lowest to highest fluences.

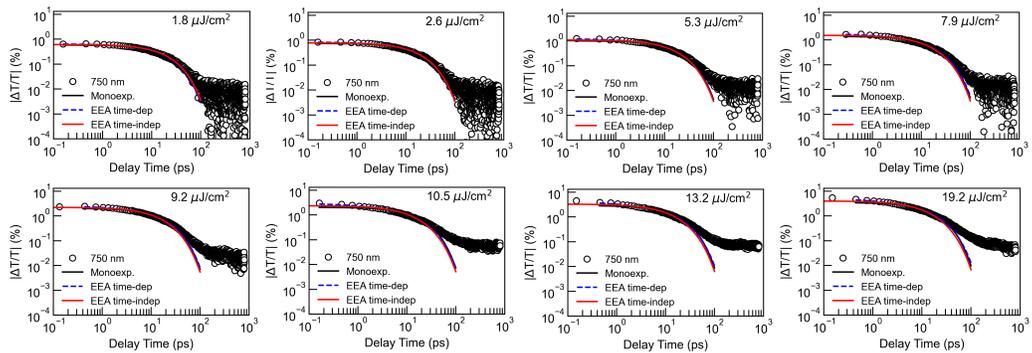

**Figure S2:** The temporal decay measured from low to high fluences (black open circles). The decay traces are fit with monoexponential decay (black solid line), bimolecular annihilation models with time-independent (red solid line), and -dependent bimolecular rates (blue dash line), respectively.

# 2 Excitation Correlation Photoluminescence Spectroscopy

## 2.1 Experimental Methods

A pump at a wavelength of 730 nm was generated using the previously mentioned Orpheus system. Subsequently, the beam was directed to a 50/50 beam splitter cube, where one of the beams was directed towards a motorized linear stage (LTS300, Thorlabs), enabling precise control over the time delay between the two pulses. Each pulse was modulated using a chopper at frequencies of 372 and 199 Hz, respectively, before being focused onto the sample using a 100 mm focal length lens. Both the total integrated response and the nonlinear component were simultaneously obtained by demodulating both the fundamental and the sum of the modulation frequencies. For photoluminescence detection (ECPL), the emitted PL underwent filtration with a 750 nm long-pass filter to eliminate the pump light, after which it was focused into a photoreceiver (New Focus 2031 PR) connected to a lock-in amplifier (HF2LI, Zurich Instruments). All the measurements are performed within a home-built vacuum chamber at ambient temperature.

## 2.2 Experimental Results and Simulations

The ECPL profiles probed for samples prepared from precursors solutions of 4, 6 and 8 g/L and their associated fits, based on Eq. 6 in the main article, are shown in Fig.S3, S4, S5, respectively. All figures show negative signals regardless the precursors' concentrations. The extracted parameters, $\gamma$ and $\beta$, are shown in Fig. 4 in the main article.

The monolecular decay rates acquired from the ECPL time-independent EEA model, are shown in Figure S6, when the setting the bound loose.

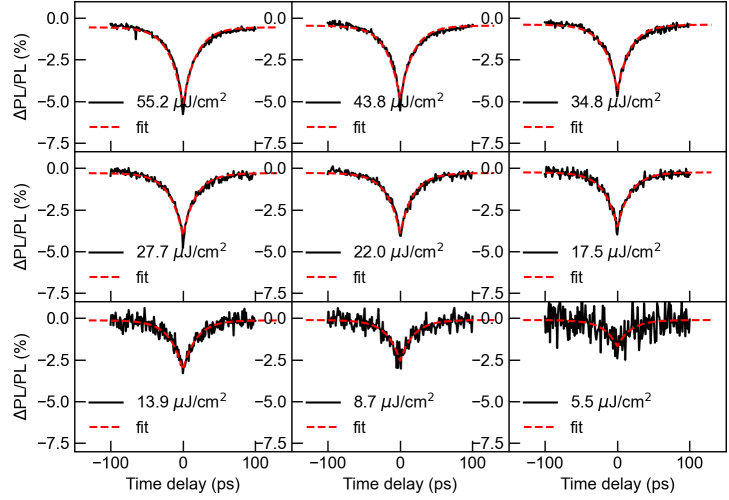

**Figure S3:** The ECPL profiles of 4 g/L (black solid line) and fits (red dash line) from the highest to the lowest fluences.

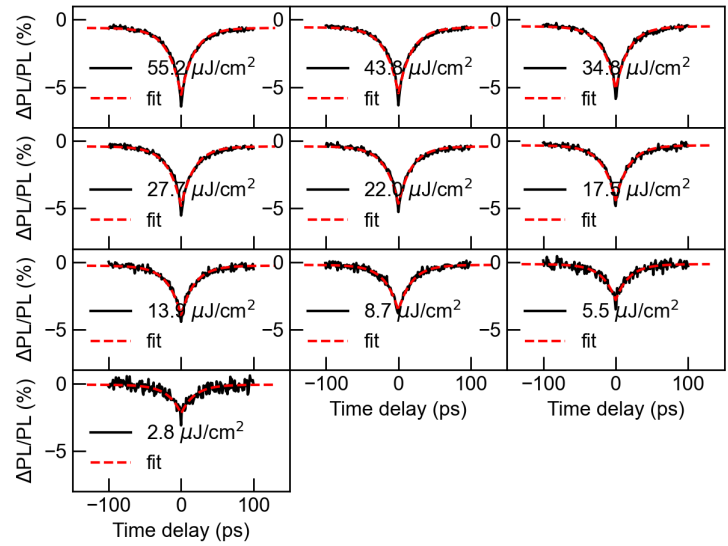

**Figure S4:** The ECPL profiles of 6 g/L (black solid line) and fits (red dash line) from the highest to the lowest fluences.

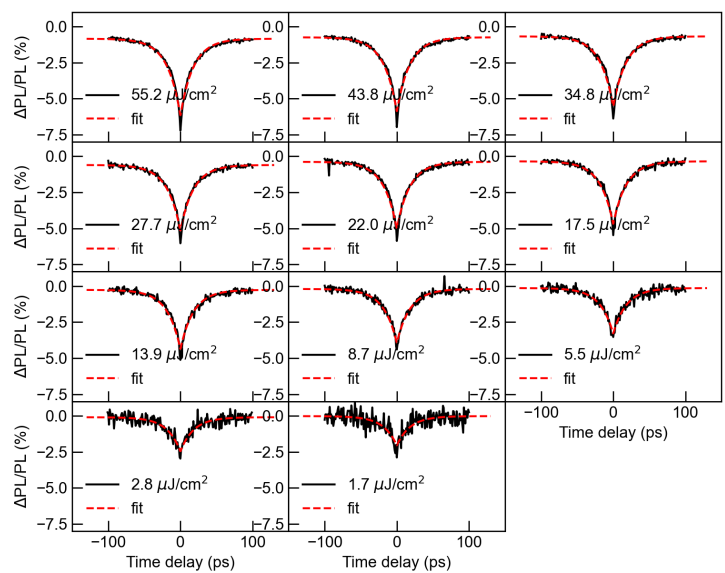

**Figure S5:** The ECPL profiles of 8 g/L (black solid line) and fits (red dash line) from the highest to the lowest fluences.

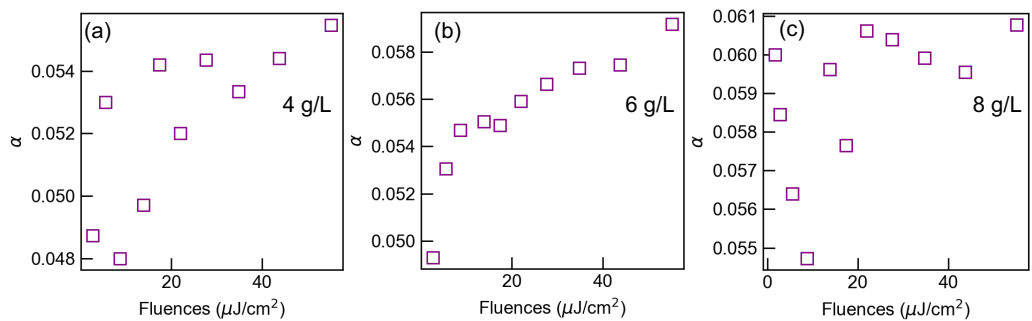

**Figure S6:** The monolecular decay rates $\alpha$ acquired from the ECPL simulations when setting the bound free for (a) 4, (b) 6, and (c) 8 g/L sample. The unit of $\alpha$ is $(\text{ps}^{-1})$.

## 2.3 Derivation of the time-integrated bimolecular model

We start from a simple bimolecular annihilation model, with monoexponential decay rate, $\alpha$ and bimolecular recombination rate, $\beta(t)$,

$$\frac{dn}{dt} = -\alpha n - \beta(t)n^2$$

$$\Rightarrow n(t) = \frac{n_0 e^{-\alpha t}}{1 + n_0 \int_0^t \beta(t) e^{-\alpha t} dt}$$

If the annihilation rate is time-independent (the diffusion is isotropic), it would give rise to,

$$n(t) = \frac{\alpha n_0}{(\alpha + n_0 \beta)e^{\alpha t} - \beta n_0}$$

For individual PL demodulated at $\omega_1$ or $\omega_2$, it will give,

$$PL_{ind} = \int_0^\infty n(t)dt = 1/\beta\{ln[(1+\gamma)]\}$$

where $\gamma \equiv \frac{n_0 \beta}{\alpha}$. Demodulate the signal at frequency $\omega_1+\omega_2$, where the $\omega_1$, $\omega_2$ are the chopping frequencies for each incident beam, then it gives the $PL_{sum}$,

$$PL_{sum} = \int_0^\tau n(t_1)dt_1 + \int_0^\infty n(t_2)dt_2$$
$$= 1/\beta\{ln[(1+\gamma)^2 - \gamma^2 e^{-\alpha\tau}]\}$$

where $\tau$ is the delay time between the two beams. The final ECPL signal will be given as,

$$\Delta PL/PL(\tau) = \frac{PL_{sum} - 2 \times PL_{ind}}{PL_{sum}}$$
$$= 1 - \frac{2ln(1+\gamma)}{ln[(1+\gamma)^2 - \gamma^2 e^{-\alpha\tau}]}$$

When $\tau=0$, it will collapse into,

$$\Delta PL/PL = 1 - \frac{2ln(1+\gamma)}{ln(2\gamma+1)}$$

The key claim that ECPL is more sensitive in probing nonlinear dynamics comes from such detection scheme. When the exciton only decays monoexponentially (i.e. $\gamma=0$), the ECPL signal will always be 0, regardless the delay time.



\end{document}